\documentclass[conference]{IEEEtran}
\IEEEoverridecommandlockouts
\usepackage{amsmath,amssymb,amsfonts}
\usepackage{algorithmic}
\usepackage{graphicx}
\usepackage{textcomp}
\usepackage{siunitx}
\usepackage{booktabs}
\usepackage{xcolor}
\usepackage[sorting=none]{biblatex}
\def\BibTeX{{\rm B\kern-.05em{\sc i\kern-.025em b}\kern-.08em
    T\kern-.1667em\lower.7ex\hbox{E}\kern-.125emX}}
\bibliography{lit.bib}
\begin{document}
\title{k-strip: A novel segmentation algorithm in k-space for the application of skull stripping}
\author{Moritz Rempe, Florian Mentzel, Kelsey L. Pomykala, Johannes Haubold, \\ Felix Nensa, Kevin Kröninger, Jan Egger, Jens Kleesiek
\thanks{This work received funding from enFaced (FWF KLI 678), enFaced 2.0 (FWF KLI 1044) and KITE (Plattform für KI-Translation Essen) from the REACT-EU initiative (https://kite.ikim.nrw/).}
\thanks{M. Rempe is with the Institute for AI in Medicine (IKIM), University Hospital Essen, Girardetstraße 2, 45131 Essen, Germany and was with the Department of Physics of the Technical University Dortmund, Otto-Hahn-Straße 4a, 44227 Dortmund, Germany (moritz.rempe@uk-essen.de).}
\thanks{F. Mentzel was with the Department of Physics of the Technical University Dortmund, Otto-Hahn-Straße 4a, 44227 Dortmund, Germany. (florian.mentzel@tu-dortmund.de}
\thanks{K. L. Pomykala is with the Institute for AI in Medicine (IKIM), University Hospital Essen, Girardetstraße 2, 45131 Essen, Germany (Kelsey.Herrmann@uk-essen.de).}
\thanks{J. Haubold, and F. Nensa are with the Institute for AI in Medicine (IKIM), University Hospital Essen, Girardetstraße 2, 45131 Essen, Germany (first.last@uk-essen.de).}
\thanks{K. Kröninger is with the Department of Physics of the Technical University Dortmund, Otto-Hahn-Straße 4a, 44227 Dortmund, Germany (kevin.kroeninger@tu-dortmund.de).}
\thanks{J. Egger is with the Institute for AI in Medicine (IKIM), University Hospital Essen, Girardetstraße 2, 45131 Essen, Germany, the Computer Algorithms for Medicine Laboratory, 8010 Graz, Austria, the Institute of Computer Graphics and Vision, Graz University of Technology, Inffeldgasse 16, 8010 Graz, Austria and the Cancer Research Center Cologne Essen (CCCE), University Medicine Essen, Hufelandstraße 55, 45147 Essen, Germany (jan.egger@uk-essen.de).}
\thanks{Corresponding author:  J. Kleesiek is with the Institute for AI in Medicine (IKIM), University Hospital Essen, Girardetstraße 2, 45131 Essen, Germany and the German Cancer Consortium (DKTK), Partner Site Essen, Hufelandstraße 55, 45147 Essen, Germany (jens.kleesiek@uk-essen.de).}}

\maketitle

\begin{abstract}

We present a novel deep learning-based skull stripping algorithm for magnetic resonance imaging (MRI) that works directly in the information rich complex valued k-space. Using three datasets from different institutions with a total of around 140000 MRI slices, we show that our network can perform skull-stripping on the raw data of MRIs while preserving the phase information which no other skull stripping algorithm is able to work with. For two of the datasets, skull stripping performed by HD-BET (Brain Extraction Tool) in the image domain is used as the ground truth, whereas the third dataset comes with per-hand annotated brain segmentations. All three datasets were very similar to the ground truth (DICE scores of 92\%-99\% and Hausdorff distances of under 5.5 pixel). Results on slices above the eye-region reach DICE scores of up to 99\%, whereas the accuracy drops in regions around the eyes and below, with partially blurred output. The output of \textit{k-strip} often has smoothed edges at the demarcation to the skull. Binary masks are created with an appropriate threshold. With this proof-of-concept study, we were able to show the feasibility of working in the k-space frequency domain, preserving phase information, with consistent results. Besides preserving valuable information for further diagnostics, this approach makes an immediate anonymization of patient data possible, already before being transformed into the image domain. Future research should be dedicated to discovering additional ways the k-space can be used for innovative image analysis and further workflows.

\end{abstract}


\section{Introduction}
\label{sec:introduction}
\IEEEPARstart{I}{mage} scanners provide more data of patients, than can actually be manually reviewed and fully annotated  in the current time-limited clinical routine. For example tumor size is often measured by maximum 2D tumor diameter, rather than a more complete tumor volume \cite{egger2012medical}, \cite{zimmermann2021ct}. Hence, there is a strong desire for automated or at least semi-automated methods, helping to process medical images, or medical information in general \cite{heiliger2022beyond}.

An important example is skull-stripping or brain extraction, the process by which the skull and non-brain tissues are removed from magnetic resonance images (MRI) \cite{kalavathi2016methods, roy2017robust, thakur2019skull}. Skull stripping is a fundamental step in neuroimage pre-processing because the accuracy of subsequent image processing, such as registration \cite{b11} or tumor segmentation \cite{menze2014multimodal}, \cite{b12}, relies on the accuracy of the skull-stripping. Not only is the process important for accuracy of further pipelines, but it is also an important step in anonymization \cite{mattern2021chemical}. Furthermore, the extracted skull information can be important for cranial implant design \cite{morais2019automated}. Unfortunately, manual removal of non-brain tissues is a complex laborious process \cite{souza2018open}, that often results in inter- and intra-rater incongruities affecting reproducibility in large scale studies \cite{thakur2019skull}. However, in recent years due to theoretical advances in the field and a rise in availability of inexpensive computing power, many deep learning-based automatic skull stripping methods have been proposed \cite{thakur2019skull}, \cite{souza2018open, hsu2020automatic, de2021automated}.

Medical images are much richer in information than what the human eye can discern \cite{avanzo2020machine}, and a lot of this untapped data is in the k-space, or the matrix of raw MRI data \cite{paschal2004k}. Quantitative imaging features, also called “radiomic features,” can provide fuller information about intensity, shape, texture, size and volume \cite{avanzo2020machine}, \cite{lambin2012radiomics}. To our knowledge, the k-space has not yet been utilized for deep learning-based segmentation. In this proof-of-concept work, we show that skull stripping is already possible in the k-space, and reconstruction of the k-space results, leads to a stripped skull in the image space. By performing the skull strip directly in k-space and not using brain extraction masks in the complex image space, we want to use possibly hidden information of the frequency domain, which are not visible in the image domain. Skull stripping directly on the raw data also enables researcher to anonymize data even before being transformed into the image domain, greatly reducing the risks of violating privacy policies. Also, this work is supposed to be a feasibility study which paves the way for future projects which can make better predictions with additional information from the k-space and advance deeper into the uncharted territory of using raw MRI data for segmentations and predictions.

\section{Related Work}
The "gold standard" for brain extraction and tissue annotation is still a manual segmentation. But because of its high time effort and the problem of inter- and intraindividual variance, leading to non-reproducible results, there has been an urgent search for methods to automate this task \cite{b13}. The currently available techniques for brain extraction can be subdivided into five groups \cite{b14}:
\begin{itemize}
 \item mathematical morphology-based methods,
 \item intensity-based methods,
 \item deformable surface-based methods,
 \item atlas-based methods,
 \item and hybrid methods.
\end{itemize}
Morphology-based methods, like by Brummer et al. \cite{b15}, use thresholds and edge-detection algorithms to distinguish between brain and non-brain tissue. The output quality depends on morphological parameters, which have to be set beforehand. Intensity-based methods extract brain tissue by solely using intensity-thresholds, based on the different intensities of brain and non-brain tissues \cite{b16}, \cite{b17}. Using curve evolution and energy functions, deformable surface-based methods perform a brain extraction by solving an optimization problem. A widely used tool applying deformable methods is \textit{BET} (Brain Extraction Tool) by the \textit{FMRIB Software Library} (FSL) \cite{b1}. Wang et al. \cite{Wang} developed an algorithm that employs a predefined atlas to create a registration mask, followed by refinements. 

Lately, with the advent of deep neural networks \cite{egger2021deep}, \cite{EGGER_MedicalMeta}, more and more algorithms have been published that use the benefits of machine learning, introducing another brain extraction method. Kleesiek et al. \cite{b2} introduced a 3D convolutional neural network (CNN), which is able to also work with pathological data, as most of the algorithms proposed before struggle when facing pathologies like tumorous tissue or abnormal anatomies. Another machine learning approach is \textit{HD-BET} by Isensee et al. \cite{b3}, which has the benefit of being trained on different sequences, like pre- and post-contrast T1, T2 and FLAIR. However, all these methods have one mayor property in common: They do the skull stripping in the image domain after a Fourier transformation has been performed.

Bassey et al. \cite{bassey2021survey} published a survey about complex-valued neural networks (CVNN), giving an insight in different works that have been done with CVNNs. These networks can take complex valued data as input, which is of importance for this work. Another field of use for these kind of networks is Quantum Computing \cite{shi2022quantum} by taking advantage of the representational capacity of complex valued numbers.

In the medical domain, deep learning approaches have mainly been used for interpolation and reconstruction on raw data. Recently, Han et al. \cite{b22} proposed an approach to interpolate missing k-space data for better reconstruction results by using a convolutional neural network and a low-rank Hankel matrix completion. Huang et al. \cite{b24} combine the task of reconstruction and segmentation, creating an end-to-end framework, which uses raw data information in the reconstruction process and attention modules, possibly preserving important image features. 

By performing post processing in the image space, one does not only risk to work on data which has lost information after transformation by using undersampling reconstruction techniques, but also neglects half of the available data - the phase information. In this work, however, a novel network is proposed, which performs skull stripping in the frequency domain, the k-space. \textit{K-strip} is a network, based on U-Net \cite{b7}, extended by residual blocks and different complex valued layers, which can take complex data as input.
\section{Material and Methods}
\subsection{Fourier Transformation}

A Fourier transformation \cite{b30} converts a 2D-signal from the time-domain into the frequency-domain or from the spatial-domain into the spatial-frequency-domain by applying \ref{eq:fourier}.
\begin{equation}
\label{eq:fourier}
    S(k_x, k_y) = \int_x\int_y s(x,y)\ \text{exp} (-i2\pi k_x x) \text{exp}\ (-i2\pi k_yy)\ dx dy
\end{equation}
The inverse form (inverse Fourier transformation) of this equation is used to transform the raw MRI data into image space. No information is lost during this process. There have been some implementations of neural networks using the Fourier domain to speed up training. Mathieu et al. \cite{b32} transform the input and the convolution kernel of a CNN into the Fourier domain. Convolutions in the spatial domain are equivalent to pointwise multiplications in the frequency domain. That way, by applying the FFT on the input data, the computational complexity and time can be reduced due to less operations. In this regard, Pratt et al. \cite{b33} were able to reduce computation time for large images in CNNs by applying Fourier pooling and convolution.

\subsection{k-space}

The \textit{k-space}, or the frequency domain, is the raw data domain of MRI images. By applying a Fourier transformation, k-space data can be transformed into the image space \cite{b4}. 

The image, on which skull stripping in today's algorithms is performed, consists of the transformed magnitude data. Raw data in MRI does not only consist of magnitude, but also phase information. This information can be used in differentiating tissue types, which do not differ in magnitude. Phase information can also be used for flow detection \cite{b5}, \cite{b6}. A more detailed look into k-space and other aspects of magnetic resonance imaging can be found in \cite{b9}.

Due to sparseness of available raw data, already transformed magnitudinal data from two of the three datasets, has been inverse Fourier transformed to create artificial k-space data. These two datasets will be used to compare our network with state-of-the-art algorithms and to show that it can hold up with them even though our network is not intended to beat them, but to solve a whole different problem. One of the three datasets comes with magnitudinal and phase data, enabling us to show the feasibility of performing skull stripping on raw data consisting of magnitude and phase. 

\subsection{Datasets}

Three datasets are used in this work, two of them provided from the University Hospital Essen.\footnote[1]{Approved by local ethics board: IRB 21-10487-BO} The first of these two datasets consists of 30000 $T_1$ (1.5 T and 3.0 T) brain MRI 2D-slices from 207 patients in image space. Some of the scans contain pathologies like glioblastomas multiforme (GBM) \cite{egger2013gbm}, giving the possibility to prove the feasibility of working with pathological data in k-space. The second dataset also comes from the University Hospital Essen and consists of 83000 susceptibility weigthed imaging (SWI) 2D-slices from 433 patients with a resolution of 2mm$^3$. This dataset also comes with the corresponding $T_2$ magnitude and phase images. To recreate real raw data, the magnitude and phase images are combined into complex valued images $z$ by scaling the phase data into the range [0; 2$\pi$] and applying 
\begin{equation}
    z = r \cdot e^{i\cdot\phi},
\end{equation}
with $r$ being the magnitude data and $\phi$ the scaled phase data. One difference of the SWI dataset to real raw data is the low-pass filtering of the phase information performed as a postprocessing step. Nevertheless we are able to show the feasibility of our algorithm on real world data with this dataset. 
The ground truth for both datasets is generated by applying the skull-stripping algorithm \textit{HD-BET} \cite{b3} on the $T_2$ images in the image domain.
The last dataset is the publicly available Neurofeedback Skull-stripped (NFBS) repository (http://preprocessed-connectomes-project.org/NFB\_skullstripped) \cite{puccio2016preprocessed}, containing 24000 fully sampled brain $T_1$ MRI scans from 125 patients with a resolution of 1mm$^3$. The dataset is already anonymized and the ground truth is created by manual refinements of an initial brain segmentation by the BEaST algorithm \cite{eskildsen2012beast}.
All slices all preprocessed by resizing them into the shape $256\times 256$ pixel and transformed into k-space via the PyTorch Fast Fourier Transformation function \textit{torch.fft} \cite{b29}.
Our network does not need further preprocessing and can work with unnormalized complex valued data. The data is augmented by randomly increasing the values by a factor in the range [0.7; 1.3] of the periphery with a random extension of the range [5; 40] in all direction of the 2D-slice. This \textit{periphery-augmentation} enables us to augment the data immediately in the complex valued k-space without the need to transform the data back into the image domain.
All three datasets are split patient-wise into training, validation and test cases by 70\%-20\%-10\%, respectively.

\subsection{Network Architecture}
\begin{figure*}[!t]
\centerline{\includegraphics[width=\textwidth]{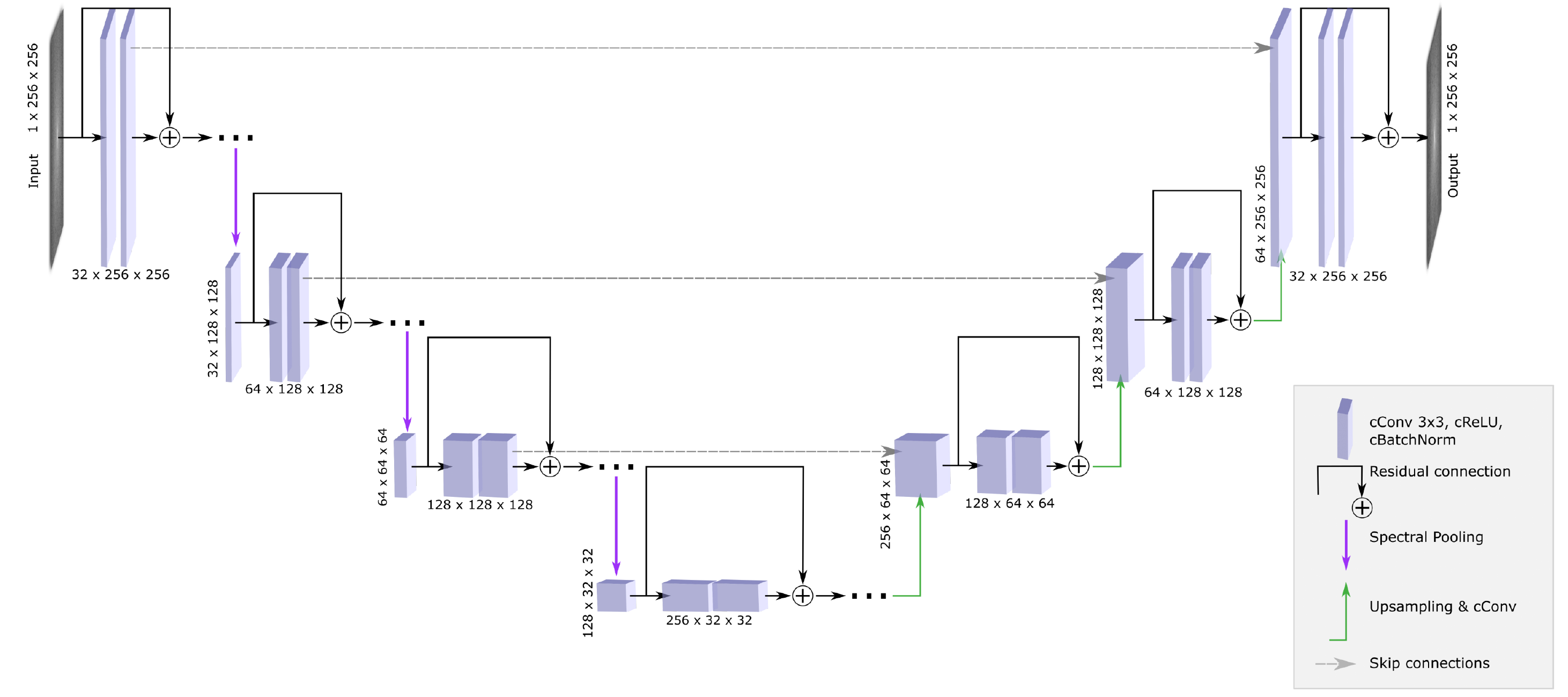}}
\caption{Network architecture of k-strip, with a down-sampling path (left branch) and an up-sampling path (right branch). The blue boxes correspond to the convolutions inside the residual blocks and their feature maps. The numbers beneath and at the side of the boxes depict the input numbers of channels, height and width. For a better gradient flow, residual connections between the double convolutions are implemented. In the downsampling branch and the bottleneck, each layer consists of four consecutive residual blocks. The upsampling branch uses nearest-neighbour-upsampling in combination with a complex valued convolution to increase the inputs dimensionality. Input and output are shown magnitudinal and logarithmically.}
\label{fig:k-strip}
\end{figure*}
K-strip is a complex valued convolutional neural network consisting, like U-Net, of a down-sampling (encoder) path and an up-sampling (decoder) path. The architecture of k-strip is shown in Fig. \ref{fig:k-strip}. It takes as input the complex valued k-space data of the MRI scan, consisting of brain tissue and skull in the frequency domain, and outputs a complex valued k-space which only consists of the brain tissue in the frequency domain. In the down-sampling path (left side of Fig. \ref{fig:k-strip}), the input is split into multiple feature channels and is down-sampled via pooling. With increasing network depth, learned features are becoming more specific, while also increasing the computational complexity due to the increasing number of feature maps. By pooling after each convolution layer, this complexity is reduced to speed up the computation time and reduce the memory usage. The initial 2D-input size is 1 $\times$ 256 $\times$ 256 pixel (number of channels $\times$ height $\times$ width) and is abstracted into feature maps via consecutive complex double convolutions inside the residual blocks to the size of 256 $\times$ 32 $\times$ 32 pixel. The input is then up-sampled again, until it reaches its initial dimensionality. 

The complex convolution blocks are wrapped inside residual blocks. Residual blocks have the benefit of allowing an easier gradient flow by using residual connections between in- and output of the convolution blocks, such that a bigger variety of information can be used by the network. In the k-strip network, in each encoder-layer the input is iterated through four residual blocks before being downsampled into the next deeper layer. 
The concept of complex convolution layers is the same as of normal convolution layers, with the difference, that instead of a simple convolution, we now work with a complex convolution matrix $W$, which acts as the kernel of size $3 \times 3$ and performs a complex convolution (\ref{eq:complex conv matrix}), as seen in \cite{b8}:
\begin{align}
\begin{split}
\label{eq:complex conv matrix}
    &\text{Complex convolution matrix: } \\
    &W = X + iY \\
    &\text{Complex data: } \\
    &d = a + ib \\
    &\text{Complex convolution: } \\
    &W * d = (X * a - Y * b) + i(X * b - Y * a)
\end{split}
\end{align}
Each \textit{cConv}-block consists of two complex convolutions following a complex ReLU (cReLU) activation function and a complex Batch normalization. After each cConv-block, the number of features is doubled.
\textit{cReLU} applies a ReLU activation on the real and the imaginary part separately and combines the output back into a complex number:
\begin{equation}
    \text{cReLU}(\{d\}) = \text{ReLU}(Re\{d\}) + i\text{ReLU}(Im\{d\}) .
\end{equation}
\begin{figure}[!t]
\centerline{\includegraphics[width=\columnwidth]{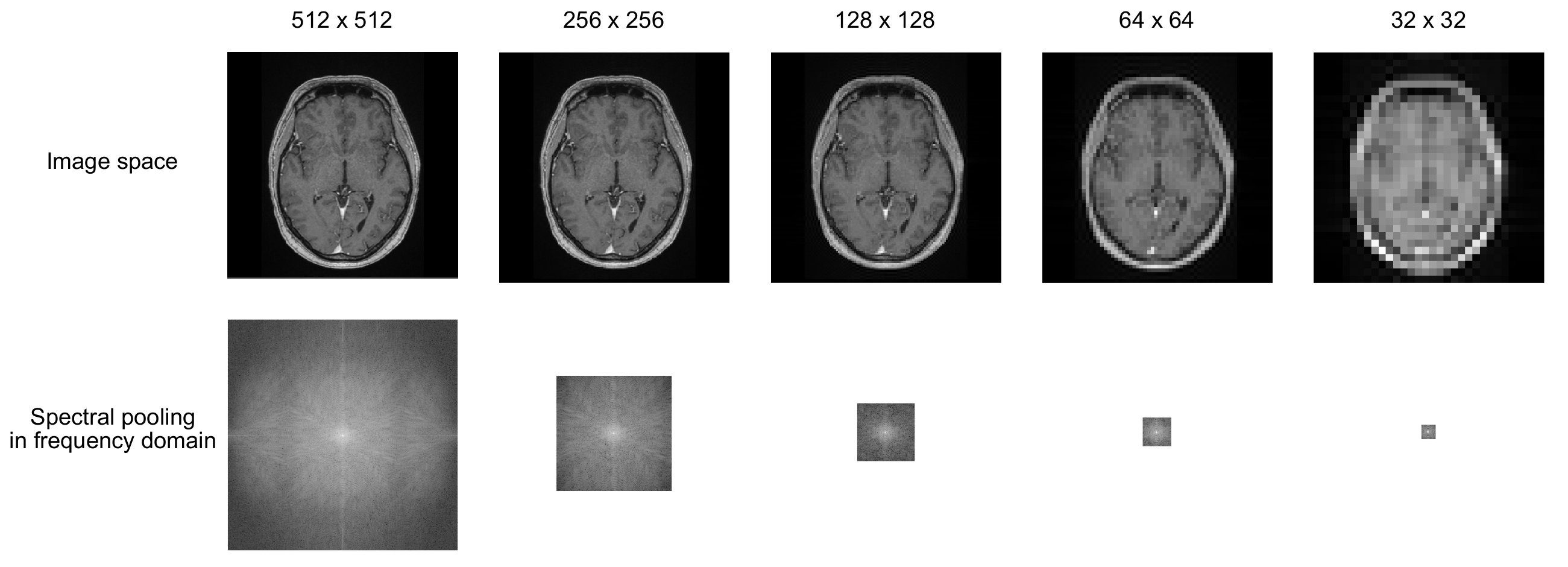}}
\caption{Spectral pooling with a pooling kernel of size two. The k-space, 
pictured logarithmic and absolute valued, is halved with every pooling 
step, which leads to a decreasing resolution. Even after reducing the 
initial k-space by 95\%, the object shape is still recognizable.}
\label{fig:spectral}
\end{figure}
By applying a complex batch normalization (\ref{eq:BatchNorm}, as seen in \cite{b28}) after each activation function, the output is normalized per batch, preventing unbalanced gradients within the network \cite{b10}:
\begin{align}
\begin{split}
\label{eq:BatchNorm}
    &\text{Complex batch normalization of $x$:} \\
    &\tilde{x} = (V)^{-\frac{1}2}(x - \text{E}[x]) \\
    & \text{Covariance matrix:} \\
    &V = \begin{pmatrix}
    V_{rr} & V_{ri} \\
    V_{ir} & V_{ii}
    \end{pmatrix} \\
    &  = \begin{pmatrix}
    \text{Cov}(Re\{x\}, Re\{x\}) & \text{Cov}(Re\{x\}, Im\{x\}) \\
    \text{Cov}(Im\{x\}, Re\{x\}) & \text{Cov}(Im\{x\}, Im\{x\})
    \end{pmatrix}
\end{split}
\end{align}
Spectral pooling \cite{b26} reduces the image size, by cutting the periphery of the k-space by half its size in each layer. Even with cutting the high frequencies and thus reducing the input size by over 95$\%$, the initial shape is still recognizable. By downsampling the input, the learned feature maps can focus on the most important structural elements, while reducing the computational complexity drastically. Spectral pooling on exemplary brain MRI images is shown in Fig. \ref{fig:spectral}.
In the upsampling (decoder) path, instead of the classical transposed convolution, we use a nearest neighbour upsampling approach combined with a complex valued convolution to increase the images dimensionality again. It showed in our experiments, that transposed convolutional layers led to vanishing gradients, hindering the ability of our network drastically. 
Skip connections provide an additional path for the gradient by skipping some layers and feeding the output directly into the next layers. That way, the problem of vanishing gradients can be avoided, as even when some gradients approach zero, they are complemented by gradients of previous layers \cite{b27}.

 K-space consists of different frequencies, each influencing many parts of the image in image space. Because of this property, a traditional assignment into two classes, as done in image space, is not possible. Therefore, our network translates the input k-space into a k-space of a skull stripped scan by adjusting the frequency values according to the principle of \textit{image-to-image} networks. That is the process of transforming an input image (here the k-space with skull) into another image (skull stripped k-space), by changing the values of the frequencies accordingly. In our case, one can also call it \textit{frequency-to-frequency}. 

\section{Experiments}
\label{sec:experiments}

\subsection{Training}

For training, the Adam optimizer \cite{kingma2014adam} with an initial learning rate of 1e-3, which is reduced by 50\% each 50th epoch, a beta coefficient of 0.99 and an epsilon value of 1e-08 is used. After each encoder convolution a dropout of 5\% is chosen. The network output is compared with the ground truth using the complex L1-loss, also known as least absolute error (LAE). The ground truth consists of the skull-stripped brains transformed into k-space. Training stops with the 150th epoch, because the validation loss does not further decrease at around this point. With a batch size of 64, the overall training time takes three days on an \textit{NVIDIA A100} GPU with 86GB of graphics memory. 

\subsection{Evaluation Metrics}

For evaluation we decided to use two metrics, one being the Dice coefficient (DSC) \cite{b34} as an overlap measure between the brain segmentation $X$ and the reference mask $Y$:
\begin{equation}
    \text{DSC} =  \frac{2|X\cap Y|}{|X| + |Y|} = \frac{2TP}{2TP + 2FP + 2FN} .
\end{equation}
The Dice coefficient takes values in the range of [0$\%$, 100$\%$], with 100$\%$ being a perfect segmentation, identical to the ground truth. 
The directed Hausdorff distance (DHD) \cite{b35} is used for evaluating maximum deviations, with a value starting at zero pixel for no deviations, increasing with wider outliers in the segmentation mask:
\begin{equation}
    \text{DHD}(X, Y) = \underset{x\in X}{\max}\: \underset{y\in Y}{\min}||x-y|| .
\end{equation}
Accuracy, sensisitivity and specificity are used as additional evaluation metrics.
All results are compared to the BrainSuite algorithm (BSE) \cite{shattuck2002brainsuite} from Shattuck and Leahy, whereas on the NFBS dataset we also compare to HD-BET and our own network with less training samples. Even though our approach is trying to solve a different problem than the compared methods (skull stripping in frequency domain and thus preserving phase information) we want to show, that it still can compete with state-of-the-art approaches in the brain regions in which phase information is desired to be preserved. All slices which consist of less than 5000 pixels of brain tissue are not considered for evaluation, which means that the lower part of the brain-stem and the top most brain slices are not evaluated. This can be justified by the fact that there are no more phase information of relevance for further workflows in these regions.
To compare k-strip with methods working in the image domain, the output needs to be Fourier transformed, followed by creating a binary mask. Due to the nature of the Fourier transformation, the output contains some frequencies which result in shadows around the skull stripped image. That is the reason for creating the binary mask. A threshold is used to classify the two classes (brain and non-brain tissue). After quantitative evaluations, the threshold is set to 1.7 times the mean value of the predicted image domain data. All values below threshold are set to zero.

\section{Results}
\begin{figure}[!t]
\centerline{\includegraphics[width=\columnwidth]{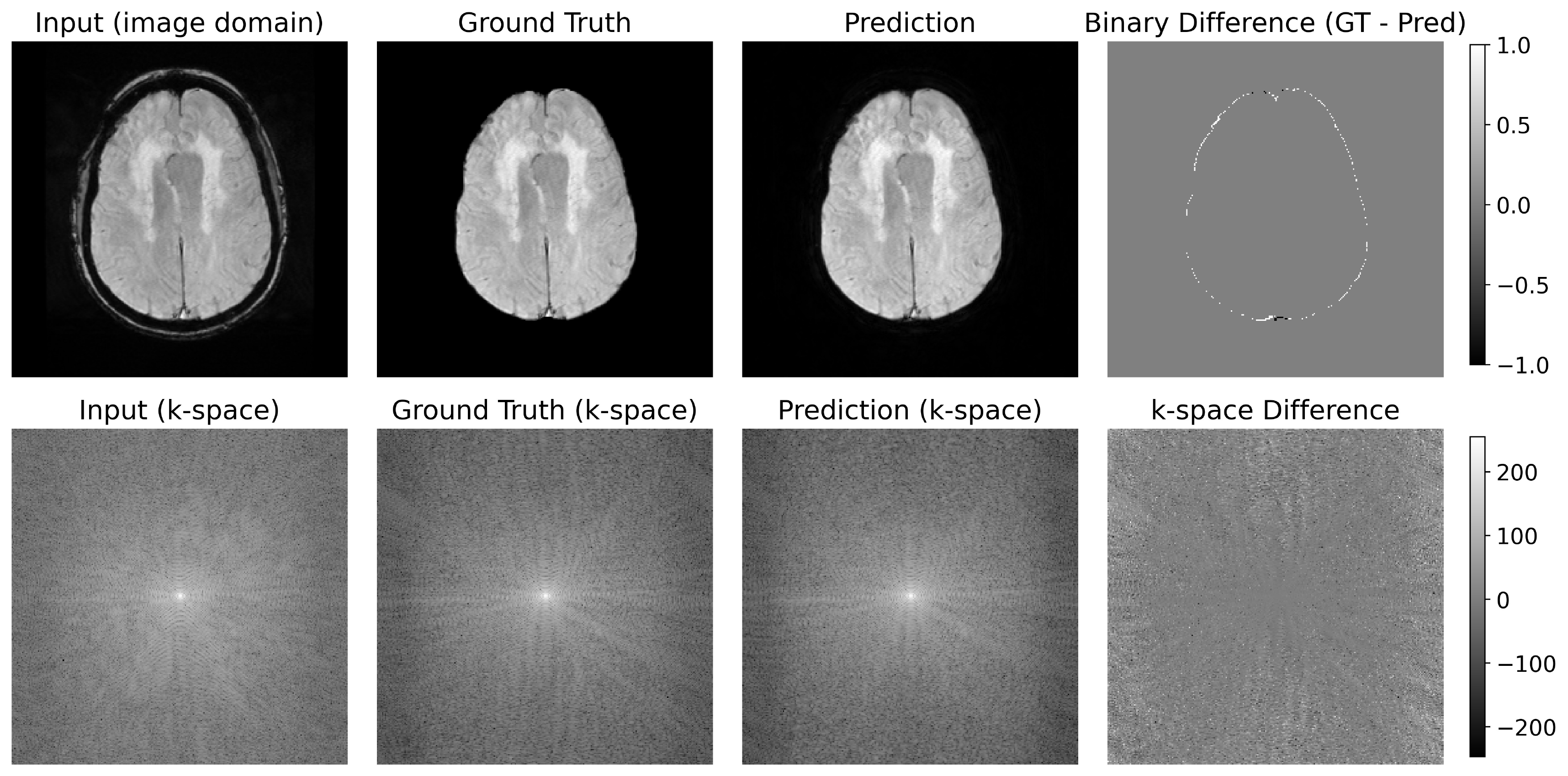}}
\caption{Magnitude output example from the SWI test set: Shown on the left side is the magnitudinal input transformed into image space (top) and the magnitudinal logarithmic k-space scaled in the range [0; 255] (bottom) (the complex valued k-space data being fed into the network). The networks prediction is transformed into image space (top) and then via threshold transformed into a binary mask. On the right side is the difference (ground truth - prediction) of the binary masks and the logarithmic magnitude of the k-spaces, respectively.}
\label{fig:swi_mag}
\end{figure}

\begin{figure}[!t]
\centerline{\includegraphics[width=\columnwidth]{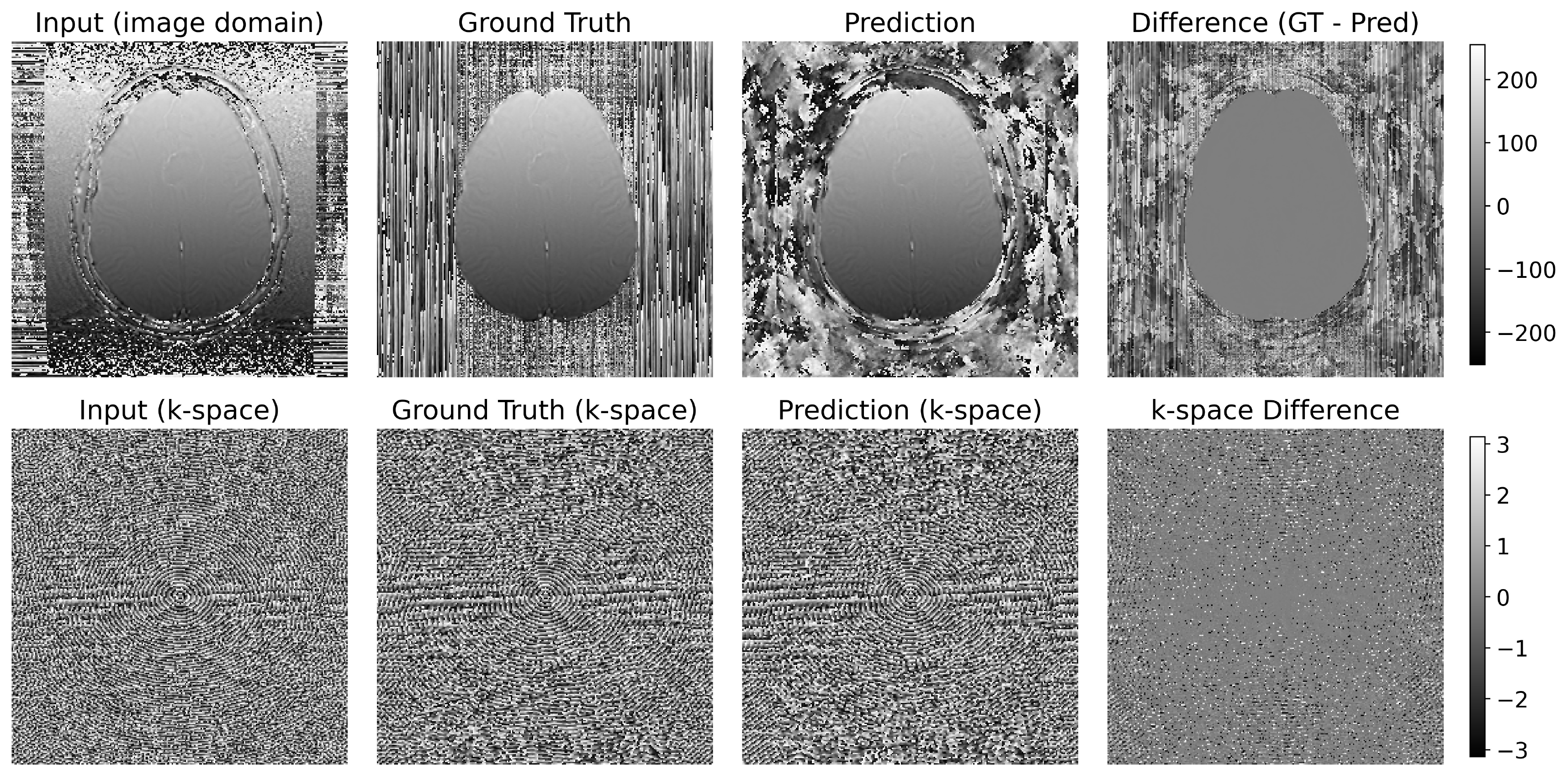}}
\caption{Corresponding phase output from the SWI test set example: Shown on the left is the 
input transformed into image space and the logarithmic angular phase
k-space (the complex valued k-space data being fed into the network) scaled in the range [$-\pi$; $\pi$].}
\label{fig:swi_pha}
\end{figure}

\begin{figure}[!t]
\centerline{\includegraphics[width=\columnwidth]{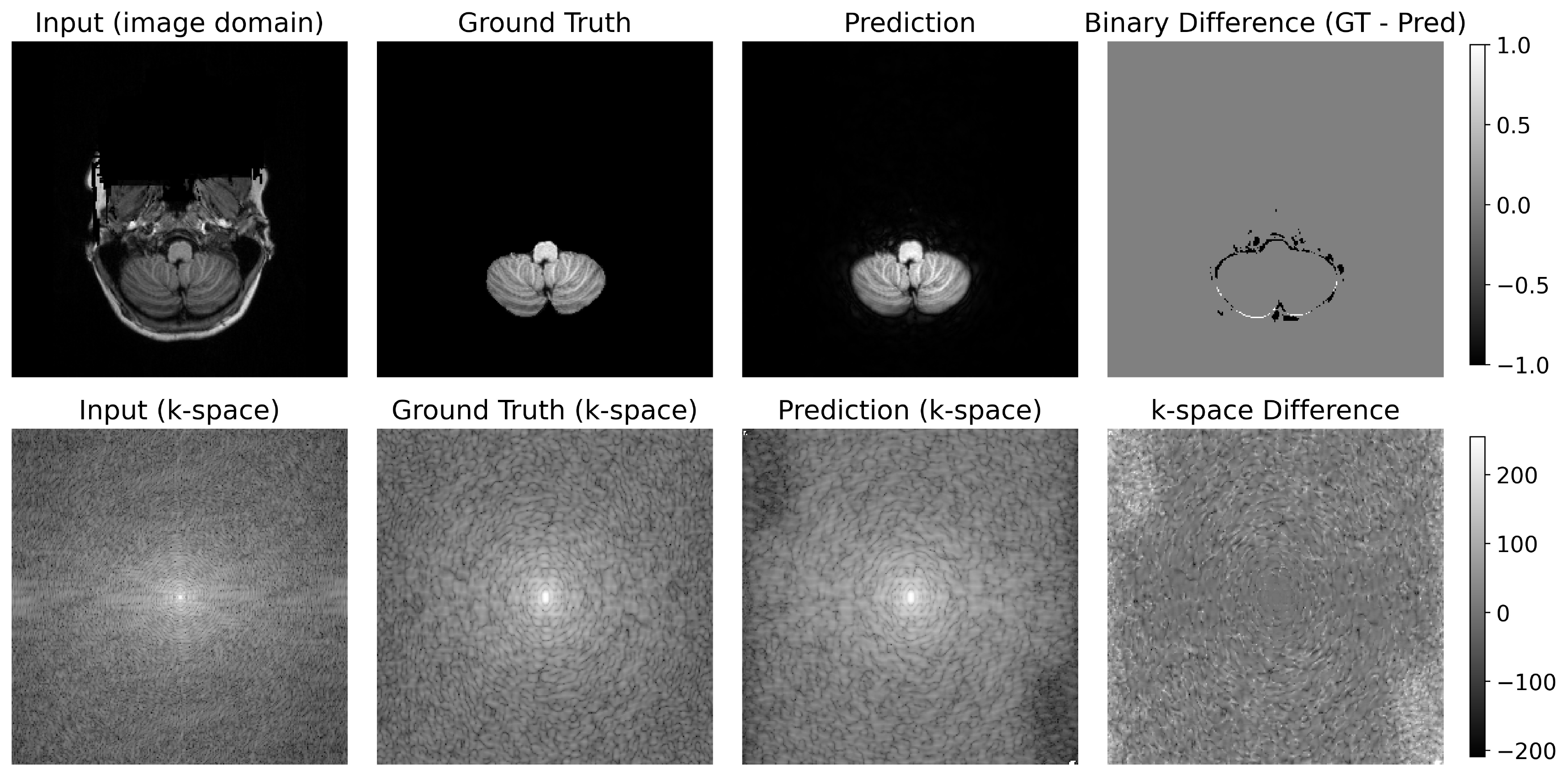}}
\caption{Output example from the Neurofeedback Skull-stripped (NFBS) test set for the brain region below the eyes: Shown on the left side is the input transformed into image space (top) and the magnitudinal logarithmic k-space scaled in the range [0; 255] (bottom) (the complex valued k-space data being fed into the network). The networks prediction is transformed into image space (top) and then via threshold transformed into a binary mask. On the right side is the difference (ground truth - prediction) of the binary masks and the logarithmic magnitude of the k-spaces, respectively.}
\label{fig:nfbs}
\end{figure}

\begin{figure}[!t]
\centerline{\includegraphics[width=\columnwidth]{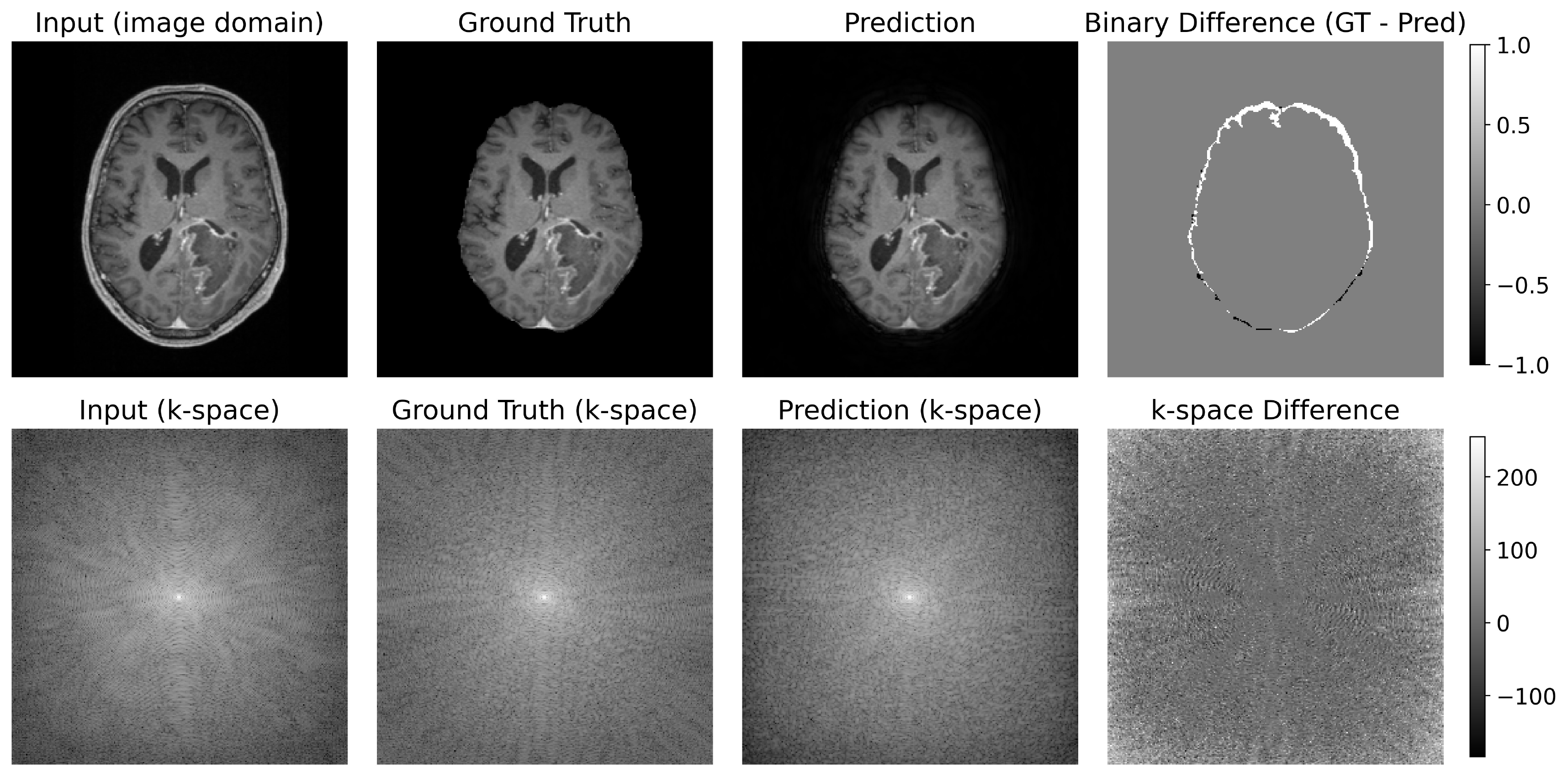}}
\caption{Output example from the glioblastoma multiforme (GBM) test set from the brain region above the eyes: Shown on the left side is the input transformed into image space (top) and the magnitudinal logarithmic k-space scaled in the range [0; 255] (bottom) (the complex valued k-space data being fed into the network). The networks prediction is transformed into image space (top) and then via threshold transformed into a binary mask. On the right side is the difference (ground truth - prediction) of the binary masks and the logarithmic magnitude of the k-spaces, respectively.}
\label{fig:gbm}
\end{figure}

Fig. \ref{fig:swi_mag} and Fig. \ref{fig:swi_pha} show the predicted magnitude and phase from a sample of the SWI test set, respectively. For visualization the k-space is shown as the logarithmic magnitude (bottom row), but the networks input consists of the complex valued k-space. To understand the input and output data, the top row shows the Fourier transformed image domain, but it has to be kept in mind, that these images are not used in the whole workflow, except for visualization and evaluation. A binary mask of the output and ground truth in image domain is calculated via the threshold method to then calculate the difference (ground truth - prediction). The phase output consists of the segmented brain and the background noise, which makes it harder to appreciate the result. But still it is obvious that the phase data of the brain is successfully segmented when looking at the difference between the ground truth and the prediction on the right side. Unlike the other figure showing the magnitudinal output, the difference is not calculated on the binary masks, but on the output and the ground truth directly to show, that no phase information is lost in the process of segmenting the raw data.
Fig. \ref{fig:nfbs} shows an exemplary output for the NFBS test set of the region below the eyes. In this region of the brain, the network still reaches Dice scores from over 95$\%$. Below this region accuracy drops, as the output gets partially blurred and the threshold method does not create proper binary masks anymore. 
A result from the GBM test set is shown in Fig. \ref{fig:gbm} from above the eyes. The network output is smoothed around the edges in comparison to the ground truth. k-Strip achieves confident results also in the presence of large pathologies as seen in this example. The network predicts the center of the k-space more accurately than the periphery, with the largest deviations from the ground truth in the far periphery which is cut out earlier due to the spectral pooling. 
The calculated evaluation metrics for all test sets and in comparison with the BSE algorithm are listed in Table \ref{tab:1}. On the NFBS test set we also compared with the HD-BET algorithm. k-Strip reaches better scores on the SWI test set, with an overall Dice score of 97.4$\%$ and a directed Hausdorff distance of 3.35 pixel. In the publicly available dataset NFBS, HD-BET achieves a slightly better Dice score and Hausdorff distance, whereas BSE is on par with k-Strip.
\begin{table*}[!t]
\caption{Scores for all three test sets (SWI, GBM, NFBS) of our proposed k-Strip network and the BSE algorithm as comparison method. On the NFBS test set, HD-BET is used as an additional comparison method. The evaluation metrics are Dice score, directed Hausdorff distance, accuracy, sensitivity and specificity, listed in this order. On the SWI dataset, k-Strip is compared with itself when trained with a different amount of training data (25000 and 70000 slices).}
\begin{center}
\def\arraystretch{1.3}
\begin{tabular}{cc|c|c|c|c|c}
\hline
\multicolumn{2}{c|}{} & {DICE ($\%$)} & {DHD (pixel)} & {Acc ($\%$)} & {Sens ($\%$)} & {Spec ($\%$)} \\
\cline{1-7} 
\multicolumn{2}{l|}{SWI} & \multicolumn{5}{c}{} \\
& BSE \cite{shattuck2002brainsuite} & 93.4 & 4.04 & 96.6 & 95.5 & 97.0 \\
& \textbf{k-Strip} (70000) & 97.4 & 3.42 & 98.9 & 97.5 & 99.5 \\
& k-Strip (25000) & 96.9 & 3.35 & 98.9 & 97.2 & 99.5 \\
\hline
\multicolumn{2}{l|}{GBM} & \multicolumn{5}{c}{} \\
& \textbf{BSE} \cite{shattuck2002brainsuite} & 94.6 & 5.03 & 97.0 & 98.5 & 94.2 \\
& k-Strip & 93.9 & 4.06 & 97.4 & 94.5 & 98.8 \\
\hline
 \multicolumn{2}{l|}{NFBS} & \multicolumn{5}{c}{} \\
 & BSE \cite{shattuck2002brainsuite} & 95.8 & 3.56 & 98.5 & 98.3 & 98.5 \\
 & \textbf{HD-BET} \cite{b3} & 96.6 & 3.21 & 98.7 & 98.4 & 99.7 \\ 
 & k-Strip & 95.7 & 3.94 & 97.5 & 97.1 & 99.1 \\
 \hline
\end{tabular}
\label{tab:1}
\end{center}
\end{table*}
k-Strip is compared with itself when trained on only 25000 slices instead of 70000 slices on the SWI dataset, to see the effect of less training data on the ability to generalize, showing that even with a significantly smaller amount of training data, our model can still produce segmentations with Dice score of over 96$\%$.
k-Strips results show, that even without the phase information it can keep up with current state-of-the-art algorithms. k-Strips specificity is very high for every test set, ranging between 98.8$\%$ and 99.9 $\%$, whereas the sensitivity is lower with values between 94.4$\%$ and 97.1$\%$, as the threshold method tends to cut out the blurry edge of the segmented brain.
Brain extraction in the region above the eyes (as in Fig. \ref{fig:swi_mag}) works very well with DSCs above 95\%. The accuracy drops in regions around the eyes and below, with partially blurred output. Poor skull stripping may also occur because of threshold problems leading to erroneous binary masks.
\begin{figure*}[!t]
\centerline{\includegraphics[width=17cm]{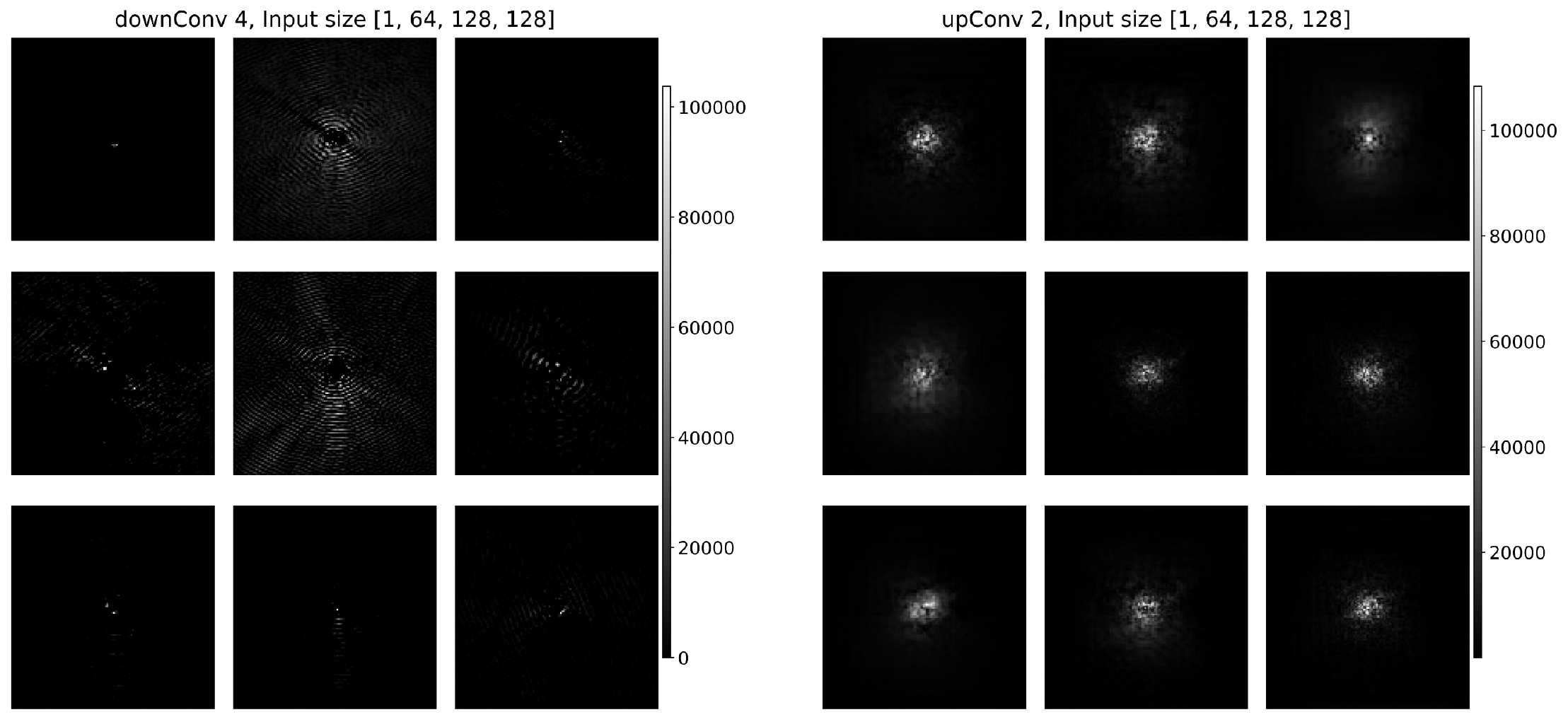}}
\caption{Magnitudinal feature maps of the fourth downsampling convolution and the second upsampling convolution. The input size of the given convolution is written in brackets with the shape [channel, batch-size, height, width], whereas the feature maps in the downsampling path specialize on k-space center and periphery individually, the convolutions in the upsampling path seem to focus on the k-space center.}
\label{fig:filter}
\end{figure*}
In Fig. \ref{fig:filter} we show exemplary feature maps of two different convolution layer. For visualization the k-space shown in the figures is magnitudinal, whereas the networks actual feature maps are complex valued. The downsampling feature maps show a specialization into the k-space center, the low frequencies corresponding to texture features, and the k-space periphery, the high frequencies corresponding to image details, individually. The upsampling (decoder) path, in contrast, mainly focuses on the k-space center, while being stretched further into the periphery in each feature map. 

When using skull stripping in the clinical routine or when processing large amounts of patient data, the inference time is one crucial factor for considering the proper algorithm. Our proposed method takes around 0.02s to segment one complex valued 2D slice, consisting of magnitude and phase information, with a shape of 256 $\times$ 256 pixel. For a volume consisting of 90 2D slices, this leads to a total inference time of around 2s. In our experiments, HD-BETs inference time for the same volume size with only the magnitude information in the image domain was 20s, whereas BSE took 18s. It has to be noted, that the inference times of HD-BET and BSE also included volume loading and saving, but this would only reduce the inference time by a few seconds, which still leads to inference times more than 10s longer than the one from our model.

\section{Discussion}
In this paper, we proposed a novel framework based on the U-Net architecture for brain extraction in k-space, in which the raw data of MRI is recorded. This proof-of-concept study shows the feasibility of working in the frequency domain and, thus, preserving phase information, while achieving consistent results. We performed training and testing on three different datasets consisting of $T_1$, $T_2$ and SWI scans, with one dataset containing real phase information. 

In regions above the eyes, the predicted output almost matches the corresponding ground truth, whereas differing sometimes in regions of the eyes and below. It has to be noted, that with the provided training data in the GBM and SWI dataset, k-strip can not outperform HD-BET, as its skull strips act as the ground truth. 
The output of k-strip shows often smoothed edges at the demarcation to the skull, the reason being the nature of the k-space and the chosen approach. As we only change the values in k-space, k-strip is not able to make exact cuts in the image domain. This may change further customization of the network. 

In regions below the eyes, in the region of the brain stem, our threshold method for creating binary masks, does not perform very well and a different method for creating binary masks in this region needs to be found in future works. 

With the center having values of magnitudes larger than the periphery, differences in the output are much more punished by the network (stronger weight-changes) than differences in the periphery. This leads to a higher deviation in the periphery of the k-space, leading to the smoothed edges of the segmentation. We might be able to resolve this in future works by implementing a logarithmic loss function which punishes deviations in the periphery stronger. 
We noticed, that HD-BET sometimes cuts out spots inside the brain, maybe due to unusual anomalies. K-strip does not seem to have problems with those cases, but does not produce as fine cuts as HD-BET or BSE does.

Reducing the size of the k-space input, using spectral pooling beforehand, would lead to faster training times, while preserving most of the details. A decision has to be made here whether the finest details provided by the k-space periphery are more important compared to faster training times, or vice versa.

Future work sees the training with real data directly from the scanner, implementing the network into more advanced workflows, for example enabling a immediate anonymization of the patient data even before being transformed into the image domain. We also want to expand the presented concept to other segmentation tasks, like tumor segmentation, using valuable phase information for more precise results, as well as further optimization of the existing network. 

\section{Conclusion}
K-strip is a novel complex valued CNN, which enables brain extraction (segmentation) on raw data of MRI scans in the k-space. To the best of our knowledge, this is the first Deep Learning-based network doing skull stripping in the k-space of MRI data. Hence, this work acts as a proof-of-concept study, while achieving comparable results as HD-BET and BSE, state-of-the-art algorithms, working in the image space. Our network makes it possible to preserve the phase information of the raw data which are discarded otherwise in the further workflow. Our approach also allows an anonymization immediately after data comes out of the scanner, even before transformation into the image domain, hence reducing the risk of privacy violation of patient data. We hope that in future, more works will investigate the possibilities and benefits of working in k-space, for example, preserving the "additional" phase information for tumor classification and tissue differentiation. 

\section{References}

\printbibliography[heading=none]

\end{document}